\documentclass[aps,superscriptaddress,showpacs,preprint]{revtex4}%

\usepackage{amsfonts}
\usepackage{amsmath}
\usepackage{amssymb}
\usepackage{graphicx}%
\usepackage{titlesec}

\titleformat*{\section}{\flushleft \bf \large}
\titleformat*{\subsection}{\flushleft \bf}
\titleformat*{\subsubsection}{\flushleft}
\begin{document}

\title{
Dissociation of H$_2$ molecule on the $\beta$-Ga$_2$O$_3$ (100)B
surface: The critical role of oxygen vacancy}

\author{Yu Yang}
\affiliation{LCP, Institute of Applied Physics and Computational
Mathematics, P.O. Box 8009, Beijing 100088, People's Republic of
China}
\author{Ping Zhang}
\thanks{To whom correspondence should be
addressed. E-mail: zhang\_ping@iapcm.ac.cn (P.Z.)}
\affiliation{LCP,
Institute of Applied Physics and Computational Mathematics, P.O. Box
8009, Beijing 100088, People's Republic of China}

\begin{abstract}

We systematically study the dissociation of H$_2$ molecules on the
$\beta$-Ga$_2$O$_3$ (100)B surface, with the influences of surface
oxygen vacancy being considered. After introducing the surface
oxygen vacancy, the nearest topmost O(I) atom becomes very active,
and hydrogen molecules become much easier to dissociate.

{\bf Keywords:} hydrogen, dissociation, oxygen vacancy,
first-principles calculation.

\end{abstract}

\pacs{68.43.Bc, 68.43.Fg, 68.43.Jk, 73.20.Hb}

\maketitle

\section{Introduction}

In ambient conditions, the stable form of gallium oxide is the
monoclinic structure of $\beta$-Ga$_2$O$_3$
\cite{Geller1960,Marezio1967,He06}, which belongs to the group of
transparent oxides with a large band gap of 4.8 eV, and exhibits
both conduction and luminescence properties
\cite{Tippins1965,Yamaga03,Yamaguchi04,Kalita08}. The
$\beta$-Ga$_2$O$_3$ is generally an $n$-type semiconductor due to
oxygen vacancies individually compensated by two electrons forming
shallow donors \cite{Lorenz1967,Cojocaru1974,Blanco05}. However, the
number of oxygen vacancies present in the material depends on the
growth atmosphere, and so the electrical character of the compound
can vary, in a tunable way, from insulating to conductive
\cite{Fleischer1992JMSL,Fleischer1992APA,Hajnal1999}. Because of its
tunable optical, magnetic and electrical properties, the
$\beta$-Ga$_2$O$_3$ has attracted much research interest for many
technological applications \cite{Blanco05}. For example, Ga$_2$O$_3$
is useful in the fabrication of masers \cite{Geller1960},
field-effect devices \cite{Li00}, switching memory devices
\cite{Aubay1993,Binet1996,Binet1998}, and transmitting ultraviolet
light \cite{Edwards1997,Wu00}. By monitoring its electronic
properties at different chemical atmosphere, the $\beta$-Ga$_2$O$_3$
can also be used as sensors to detect hydrogen gases
\cite{Fleischer1992SAB,Pohle00,Weh01,Bermudez06}. To detect hydrogen
molecules, it is revealed that oxygen vacancies are needed on the
Ga$_2$O$_3$ surface to explain the stretching frequencies of Ga-H
bonds observed in the infrared spectroscopy experiment
\cite{Gonzalez05}. However, the special mechanism for oxygen
vacancies to help dissociate hydrogen molecules has not been
discussed yet.

In addition, the H$_2$-Ga$_2$O$_3$ interaction is also important to
some catalytic reactions. Catalysts containing supported gallium are
known to be active in light alkane dehydrogenation and aromatization
\cite{Gonzalez05,Ono1992,Carli1993,Takahar04}. Recently, it is
highlighted that $\beta$-Ga$_2$O$_3$ alone is able to dissociate
hydrogen molecules at temperatures higher than 500 K and then
hydrogenate adsorbed CO$_2$ stepwise from formate to methoxy groups
\cite{Collins04}. Although the oxidation state of gallium ions has
been of a matter of debate in catalytic reactions
\cite{Meriaudeau1991,Collins02}, the role of the geometric
environment or surface coordinations has not been emphasized.

There are two types of coordination for Ga ions in the
$\beta$-Ga$_2$O$_3$, namely tetrahedral and octahedral, and referred
to as Ga(I) and Ga(II). There are also three types of oxygen ions,
referred to as O(I), O(II) and O(III). Oxygen O(I) and O(II) lie
respectively in and out of the symmetry plane (see Fig. 1(a)). They
are both in threefold coordination while O(III) is in fourfold
coordination \cite{Yamaguchi04,Blanco05}. The most frequent cleavage
plane for $\beta$-Ga$_2$O$_3$ is the (100) plane
\cite{Bermudez06,Gonzalez05,Lovejoy09}. A recent theoretical study
of the (100) surface presents two possible surface terminations, A
and B \cite{Bermudez06,Lovejoy09}. Termination A is characterized by
rows of oxygen, and termination B by rows of nearest neighbor
gallium and oxygen. In our study, we mainly focus on the surface
relaxations for the stoichiometric and oxygen vacancy included
$\beta$-Ga$_2$O$_3$ (100)B surfaces, and the dissociation
prorperties of hydrogen molecules on them. The rest of the paper is
organized as follows. In Sec. II, we describe our first-principles
calculation methods; in Sec. III, we present in detail our
calculated results, including the comparisons of the electronic
structures of the $\beta$-Ga$_2$O$_3$ (100)B surface including
oxygen vacancy or not, and the dissociation of hydrogen molecules on
the surfaces; Finally, the conclusions are given in Sec. IV.

\section{Computational method}

Our calculations are performed within density functional theory
(DFT) using the Vienna {\it ab-initio} simulation package (VASP)
\cite{VASP}. The PW91 \cite{PW91} generalized gradient approximation
and the projector-augmented wave potential \cite{PAW} are employed
to describe the exchange-correlation energy and the electron-ion
interaction, respectively. The cutoff energy for the plane wave
expansion is set to 400 eV, which is large enough to make the error
from calculations of the adsorption energy below 0.01 eV. The
stoichiometric and oxygen vacancy included $\beta$-Ga$_2$O$_3$
(100)B surfaces are modeled by a slab composed of 30 atomic layers
and a vacuum region of 20 \AA. The $1\times2$ supercell with the
size of 5.93 \AA~ $\times$ 6.11 \AA~ is used to study the H$_2$
adsorption. Our test calculations have shown that this supercell is
sufficiently large to avoid the interaction between adjacent
hydrogen molecules. Integration over the Brillouin zone is done
using the Monkhorst-Pack scheme \cite{Monkhorst} with
$7\times7\times1$ point grids. A Fermi broadening \cite{Weinert1992}
of 0.05 eV is chosen to smear the occupation of the bands around the
Fermi energy (E$_f$) by a finite-$T$ Fermi function and
extrapolating to $T=0$ K. During geometry optimizations, the bottom
10 layers of the $\beta$-Ga$_2$O$_3$ (100)B surface are fixed while
other Ga and O atoms are free to relax until the forces on them are
less than 0.01 eV/\AA. The calculation of the potential energy
surface for molecular H$_2$ is interpolated to 350 points with
different bond length ($d_{\rm H-H}$) and molecular height ($h_{\rm
H_2}$) of H$_{2}$ at each surface site. The calculated lattice
constants of the $\beta$-Ga$_2$O$_3$ crystal are $a$=12.45 \AA,
$b$=3.05 \AA, and $c$=5.93 \AA, and $\theta$=103.9$^{\circ}$, in
good agreement with the experimental and theoretically calculated
values \cite{Blanco05,Bermudez06}. The calculated bond length of the
free H$_2$ molecule is 0.75 \AA, also in good agreement with the
experimental value of 0.74 \AA~ \cite{Huber1979}.

\section{Results and discussion}

First, we do geometry optimizations for the stoichiometric
$\beta$-Ga$_2$O$_3$ (100) surface. There are two different types of
(100) surface terminations, respectively called as the
$\beta$-Ga$_2$O$_3$ (100)A and (100)B surfaces
\cite{Bermudez06,Lovejoy09}. The (100)A surface is formed by
cleaving the Ga-O(II) bonds above the O(II) atoms while leaving the
below Ga-O(II) bonds intact, and terminated in rows of O(II)'s lying
along the [010] direction. The (100)B surface is formed by cleaving
the Ga-O(II) bonds below the O(II) atoms, and terminated in
nearest-neighbor rows of Ga(II) and O(III) atoms, each singly
unsaturated. After geometry optimizations, we find that the topmost
O(I) atoms move out on the $\beta$-Ga$_2$O$_3$ (100)B surface, and
form O(I)-Ga(II)-O(II) rows as shown in Fig. 1(a). The calculated
surface energies for the relaxed $\beta$-Ga$_2$O$_3$ (100)A and
(100)B surfaces are respectively 1.19 and 0.76 J$\cdot$m$^{-2}$,
according well with other {\it ab initio} results of 1.13 and 0.68
J$\cdot$m$^{-2}$ \cite{Bermudez06}. The surface energy values
indicate that the $\beta$-Ga$_2$O$_3$ (100)B surface is much more
stable than the $\beta$-Ga$_2$O$_3$ (100)A surface.

There are two kinds of oxygen vacancies on the relaxed
$\beta$-Ga$_2$O$_3$ (100)B surface, respectively the O(I) and O(III)
vacancies. After geometry optimizations, we find that the
$\beta$-Ga$_2$O$_3$ (100)B surface with the O(III) vacancy is 0.56
eV lower in total energy than with the O(I) vacancy. The surface
geometry of the $\beta$-Ga$_2$O$_3$ (100)B surface with an O(III)
vacancy (Ov-surface) is shown in Fig. 1(c), where the possible
adsorption sites near the O(III) vacancy are depicted. On the
Ov-surface, the O(I) atom has a relatively large distortion. It
moves 0.28 \AA~ along the [00$\bar{1}$] direction during surface
relaxation.

Besides of introducing geometric distortions, the oxygen vacancy
also changes the surface electronic structures. Figure 2 shows the
surface charge density of the stoichiometric $\beta$-Ga$_2$O$_3$
(100)B surface and the Ov-surface. The contour map shows the charge
density distribution in the plane that is parallel to and 1.50 \AA~
away from the $\beta$-Ga$_2$O$_3$ (100)B surface. And the values are
in the unit of $e$/\AA$^3$. It is clearly shown that the O(I) atom
nearest to the O(III) vacancy has more electron distribution than on
the stoichiometric $\beta$-Ga$_2$O$_3$ (100)B surface, and so does
the nearest O(III) atoms. Using the Bader topological analysis
\cite{Bader} for the charge density, it is found that the nearest
O(I) atom gets 0.28 more electrons on the Ov-surface than on the
stoichiometric $\beta$-Ga$_2$O$_3$ (100)B surface, it is also found
that the two nearest Ga(II) atoms and the nearest Ga(I) atom
respectively get 0.61, 0.62 and 0.34 more electrons on the
Ov-surface. These extra electrons are because that the vacancy of a
surface O(III) atom supplies 2 more electrons to its nearby atoms.
And the other 0.15 electrons is around the other surface oxygen
atoms. The extra electrons around the surface Ga atoms are easy to
form Ga-H bonds when hydrogen atoms are introduced.

The projected density of states (PDOS) for the nearest surface O(I)
atom on the Ov-surface is shown in Fig. 3, together with the PDOS
for a surface O(I) atom on the stoichiometric $\beta$-Ga$_2$O$_3$
(100)B surface. From detailed wavefunction analysis, we find that
there are strong $s$-$p$ hybridizations in the electronic states for
Ga(I) and Ga(II) atoms on the $\beta$-Ga$_2$O$_3$ (100)B surface,
but the electronic states of surface O(I) and O(III) atoms do not
hybridize at all. In the PDOS of an O(I) atom on the stoichiometric
$\beta$-Ga$_2$O$_3$ (100)B surface, the $s$ states lie around 15 eV
below the $p$ states, and the $p$ electronic states contribute the
majority part of the electronic states around the Fermi energy. As
shown in Fig. 3, in the PDOS of the nearest O(I) atom on the
Ov-surface, there is a new peak (peak 1) below the Fermi energy,
which represents for the electronic states of the 0.28 extra
electrons due to the O(III) vacancy. Since electrons near the Fermi
energy are easier to join in electronic hybridizations, the extra
electrons around the surface O(I) atom make it more active than on
the stoichiometric Ga$_2$O$_3$ surface. Moreover, it is found from
wavefunction analysis that the electronic states of peak 1 tend to
occupy the oxygen $p_z$ orbital. These facts indicate that the
electronic states around the O(I) atoms are very easy to interact
with a parallel hydrogen molecule. As we will see later, hydrogen
molecules adsorbing around the surface O(I) atom on the Ov-surface
do dissociate more easily.

After analyzing the surface electronic structures, we then study the
adsorption and dissociation of hydrogen molecules, both on the
stoichiometric $\beta$-Ga$_2$O$_3$ (100)B surface and the
Ov-surface. As shown in Fig. 1(b), there are five high symmetry
sites on the relaxed $\beta$-Ga$_2$O$_3$ (100)B surface,
respectively on top of the Ga(I), Ga(II) and oxygen atoms, and in
the middle of two Ga(I) atoms and two oxygen atoms. In following
discussions, we call them as the TG1, TG2, TO, BG and BO sites. At
each surface site, there are three possible orientations for an
adsorbing hydrogen molecule, respectively along the $x$ ([001]), $y$
([010]) and $z$ ([100]) directions. After employing these notations,
we will use TG1-$x,y,z$, TG2-$x,y,z$, TO-$x,y,z$, BG-$x,y,z$, and
BO-$x,y,z$ to represent the considered adsorption channels for
hydrogen molecules. While studying the adsorption and dissociation
of hydrogen molecules on the Ov-surface, we choose TO to be the site
directly on top of the O(III) vacancy, and TG1 and TG2 sites to be
the sites on top of the nearest Ga(I) and Ga(II) atoms.

To study the adsorption properties of H$_2$ molecules, we do
geometry optimizations after initially putting an H$_2$ molecule
4.00 \AA~ away from the stoichiometric $\beta$-Ga$_2$O$_3$ (100)B
surface and the Ov-surface along all the above mentioned high
symmetry channels. During the geometry optimizations, the bottom 10
layers of the $\beta$-Ga$_2$O$_3$ (100)B surface are fixed while
other H, Ga and O atoms are free to relax until the forces on them
are less than 0.01 eV/\AA. From these relaxation calculations, we
find that there are no chemisorption states for H$_2$ molecules on
both the stoichiometric $\beta$-Ga$_2$O$_3$ (100)B surface and the
Ov-surface. All the H$_2$ molecules will finally evolve into the
physisorption states, with the adsorption energies of several tens
of meV.

We calculate the two-dimensional (2D) potential energy surface (PES)
cuts to evaluate the dissociation properties of hydrogen molecules.
Along each adsorption channel, $h_{\rm H_2}$ ranges between 0.40
\AA~ and 4.00 \AA, and $d_{\rm H-H}$ ranges between 0.45 \AA~ and
2.40 \AA. From the calculated 2D PES cuts for H$_2$ on the
stoichiometric $\beta$-Ga$_2$O$_3$ (100)B surface, we see no direct
dissociation. As examples, the 2D PES cuts along the TG1-$x,y$,
TG2-$x$ and TO-$y$ channels are listed in Figs. 4(a)-(d). As shown
in Figs. 4(a)-(c), on the stoichiometric $\beta$-Ga$_2$O$_3$ (100)B
surface, hydrogen molecules are very hard to get close to surface Ga
atoms. The energy barrier for a hydrogen molecule in its molecular
length of 0.75 \AA~ to be 1.00 \AA~ from the surface Ga(I) and
Ga(II) atoms is over than 12 eV. However, hydrogen molecules might
be easier to get close to the O atoms on the $\beta$-Ga$_2$O$_3$
(100)B surface. The energy barrier for a hydrogen molecule in its
molecular length to be 1.00 \AA~ from the surface O(I) atom is 5.2
eV, as shown in Fig. 4(d). In total, hydrogen molecules can not
dissociate on the stoichiometric $\beta$-Ga$_2$O$_3$ (100)B surface.
Even the energy barriers for hydrogen to get closer to the surface
is huge. These results indicate the inactivity of the stoichiometric
$\beta$-Ga$_2$O$_3$ (100)B surface to hydrogen molecules.

Similar to what we do for the 2D PES cuts of hydrogen molecules on
the stoichiometric $\beta$-Ga$_2$O$_3$ (100)B surface, we also
calculate the 2D PES cuts of hydrogen molecules on the Ov-surface.
To make a comparison, we list the 2D PES cuts along the TG1-$x,y$,
TG2-$x$ and TO-$y$ channels in Figs. 5(a)-(d). One can see from
Figs. 5(a)-(c) that the adsorption of hydrogen molecules does not
change too much on top of Ga(I) and Ga(II) atoms on the Ov-surface,
comparing with that on the stoichiometric $\beta$-Ga$_2$O$_3$ (100)B
surface. However, the adsorption of hydrogen molecules change a lot
on top of the O(I) atom. As shown in Fig. 5(d), there is a clear
dissociative adsorption state for hydrogen on the Ov-surface, in
which the bond length and molecular height of H$_2$ are respectively
1.80 and 0.50 \AA. The dissociation energy barrier for a hydrogen
molecule to reach this adsorption state is 2.97 eV, which is much
smaller than the approaching energy barrier on the stoichiometric
$\beta$-Ga$_2$O$_3$ (100)B surface, indicating a big enhancement of
the surface reactivity to hydrogen molecules. The PES results also
approves our previous hypothesis that due to the extra electrons
from the surface O(III) vacancy, the nearby O(I) atom are easier to
interact with hydrogen molecules. We also notice that the
dissociation energy barrier of hydrogen molecules is still very
large. But more importantly, we point out that surface oxygen
vacancies can greatly enhance the activity of the
$\beta$-Ga$_2$O$_3$ (100)B surface to interact with hydrogen. Given
more surface oxygen vacancies, and an environment with larger
hydrogen enthalpy, hydrogen molecules can be further easier to
dissociate on the $\beta$-Ga$_2$O$_3$ surface.

\section{Conclusion}

In summary, we have systematically studied the electronic structures
of the stoichiometric and oxygen vacancy included
$\beta$-Ga$_2$O$_3$ (100)B surfaces, and the dissociation properties
of hydrogen molecules on them. It is found that due to the oxygen
vacancy, the neighboring surface Ga and O atoms get some extra
electrons than on the stoichiometric Ga$_2$O$_3$ surface. The extra
electrons around the surface O(I) atom are very near to the Fermi
energy, and distribute on the $p_z$ orbital, which are thus very
active to hydrogen molecules. By calculating the 2D PES cuts for
hydrogen molecules on the stoichiometric and oxygen vacancy included
$\beta$-Ga$_2$O$_3$ (100)B surfaces, we find that hydrogen molecules
are very hard to get close to the stoichiometric Ga$_2$O$_3$
surface, but much easier to dissociate on the oxygen vacancy
included Ga$_2$O$_3$ surface. Our results indicate the enhancement
of surface activity to hydrogen molecules by introducing surface
oxygen vacancies.

\begin{acknowledgments}
This work was supported by the NSFC under grants No. 10904004, and
60776063.
\end{acknowledgments}

\clearpage

\noindent\textbf{List of captions} \\

\noindent\textbf{Fig.1}~~~ (color online) The $\beta$-Ga$_2$O$_3$
(100)B surface viewed along the (a) [010] and (b) [100] directions,
and (c) the $\beta$-Ga$_2$O$_3$ (100)B surface with an O(III)
vacancy viewed along the [100] direction. Grey and red balls
respectively represent Ga and O atoms. The black squares in (b) and
(c) represent the surface unit cells in our calculations.\\

\noindent\textbf{Fig.2}~~~ (Color online) Contour map of the charge
density distribution on the (a) stoichiometric and (b) oxygen
vacancy included $\beta$-Ga$_2$O$_3$ (100)B surfaces. The mapping
plane is 1.5 \AA~ from the Ga$_2$O$_3$ (100)B surface. The back
squares in
dashed lines represent the surface unit cells.\\

\noindent\textbf{Fig.3}~~~ (Color online) The projected density of
states for the O(I) atom on the stoichiometric and oxygen vacancy
included $\beta$-Ga$_2$O$_3$ (100)B surfaces. Fermi energy is set to
be the
energy zero.\\

\noindent\textbf{Fig.4}~~~ The 2D PES cuts for the adsorption of
hydrogen molecules along the (a) TG1-$x$, (b) TG1-$y$, (c) TG2-$x$
and (d) TO-$y$ channels on the stoichiometric $\beta$-Ga$_2$O$_3$
(100)B surface. The total energy of a free H$_2$ molecule plus that
of the
$\beta$-Ga$_2$O$_3$ surface is set to be the energy zero.\\

\noindent\textbf{Fig.5}~~~ The 2D PES cuts for the adsorption of
hydrogen molecules along the (a) TG1-$x$, (b) TG1-$y$, (c) TG2-$x$
and (d) TO-$y$ channels on the oxygen vacancy included
$\beta$-Ga$_2$O$_3$ (100)B surface. The total energy of a free H$_2$
molecule plus that of the oxygen vacancy included
$\beta$-Ga$_2$O$_3$ surface is set to be the energy zero.\\

\clearpage

\begin{figure}
\includegraphics[width=1.0\textwidth]{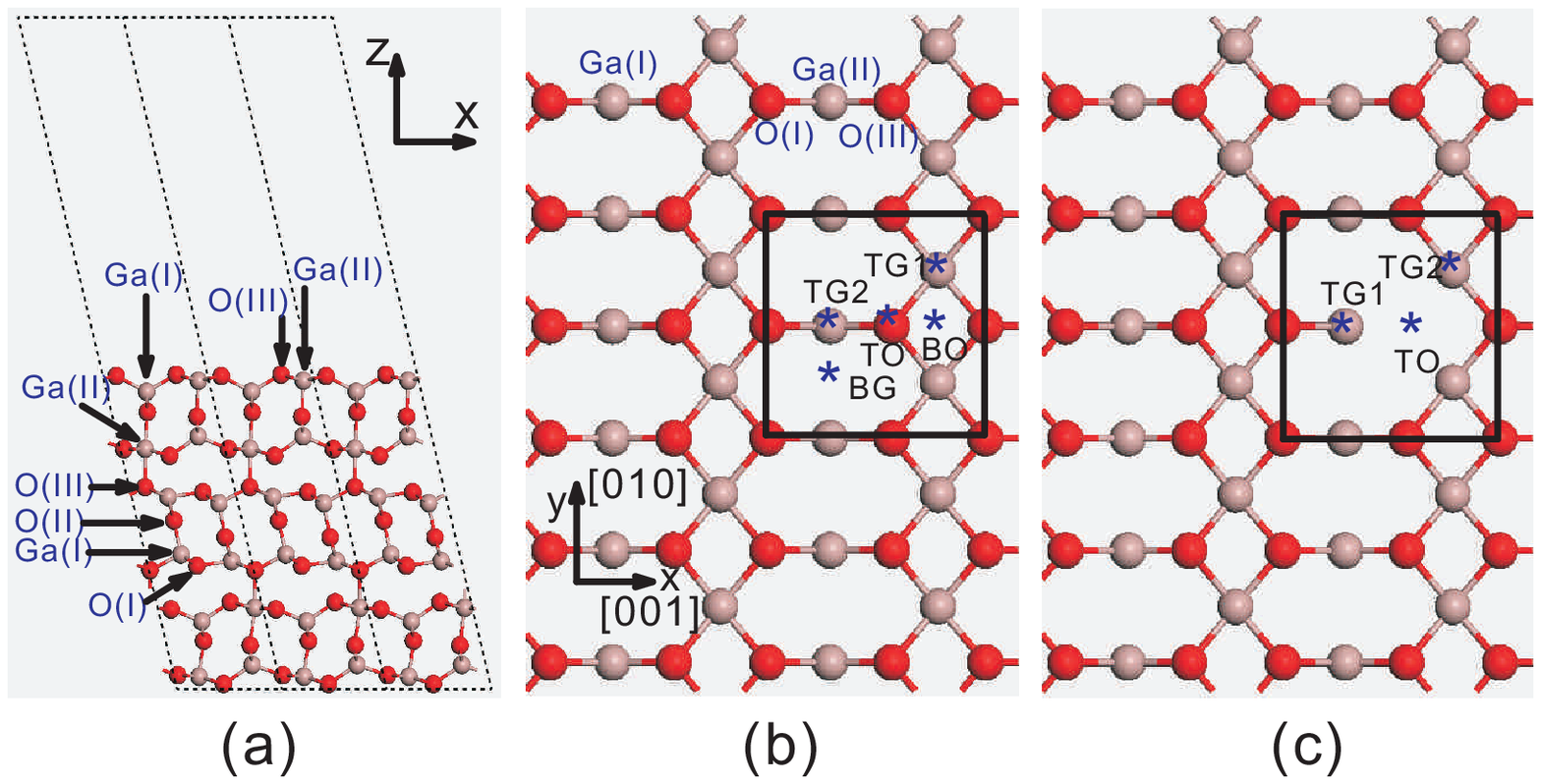}
\caption{\label{fig:fig1}}
\end{figure}
\clearpage
\begin{figure}
\includegraphics[width=1.0\textwidth]{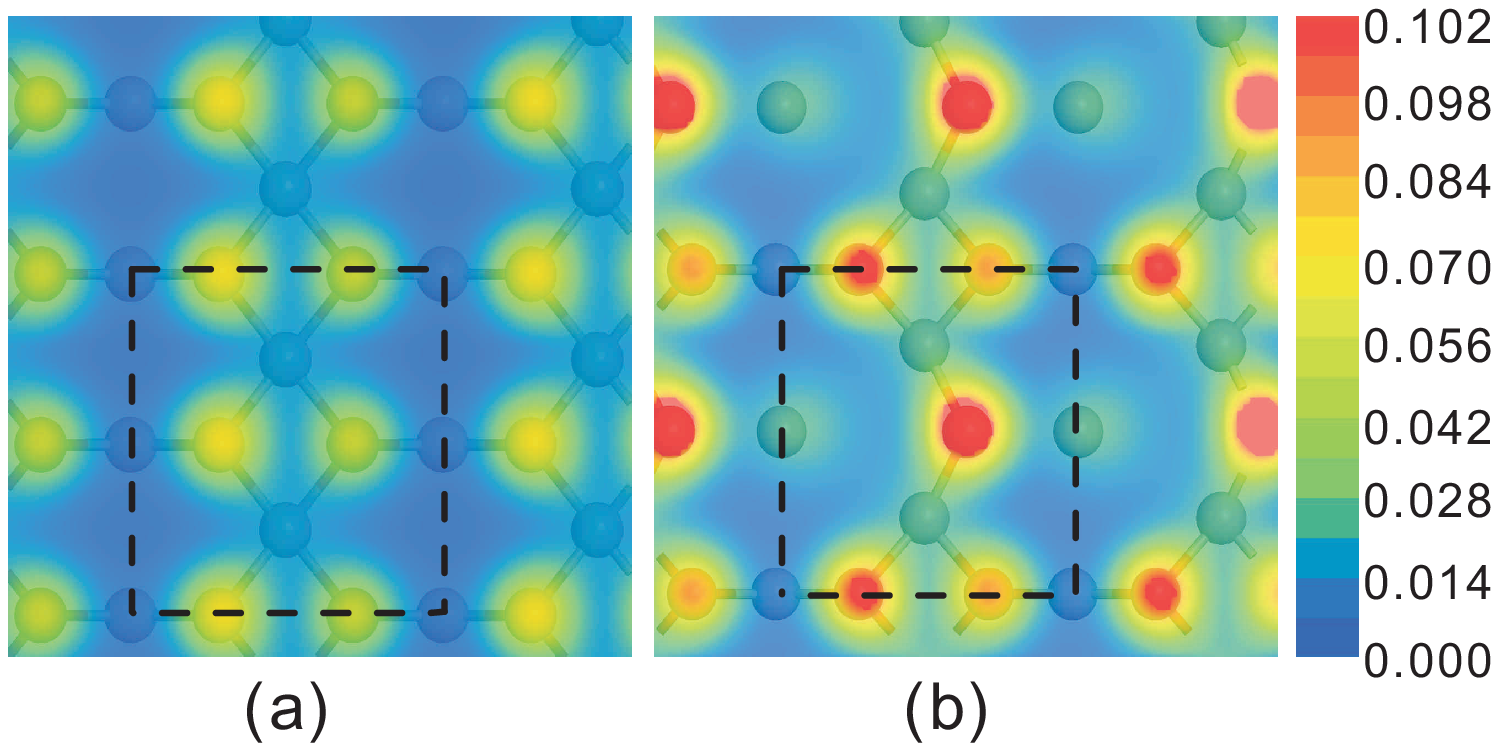}
\caption{\label{fig:fig2}}
\end{figure}
\clearpage
\begin{figure}
\includegraphics[width=1.0\textwidth]{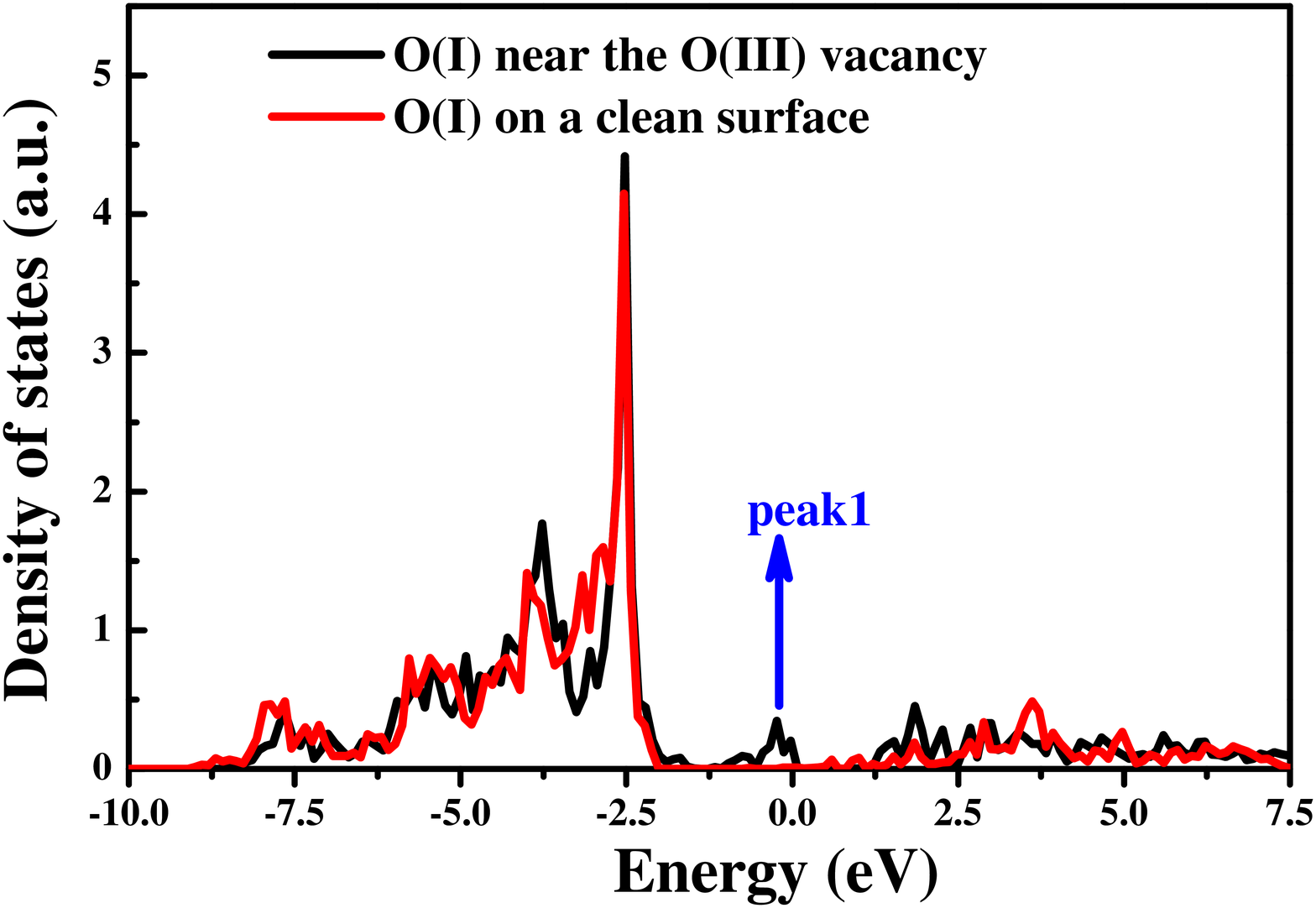}
\caption{\label{fig:fig3}}
\end{figure}
\clearpage
\begin{figure}
\includegraphics[width=1.0\textwidth]{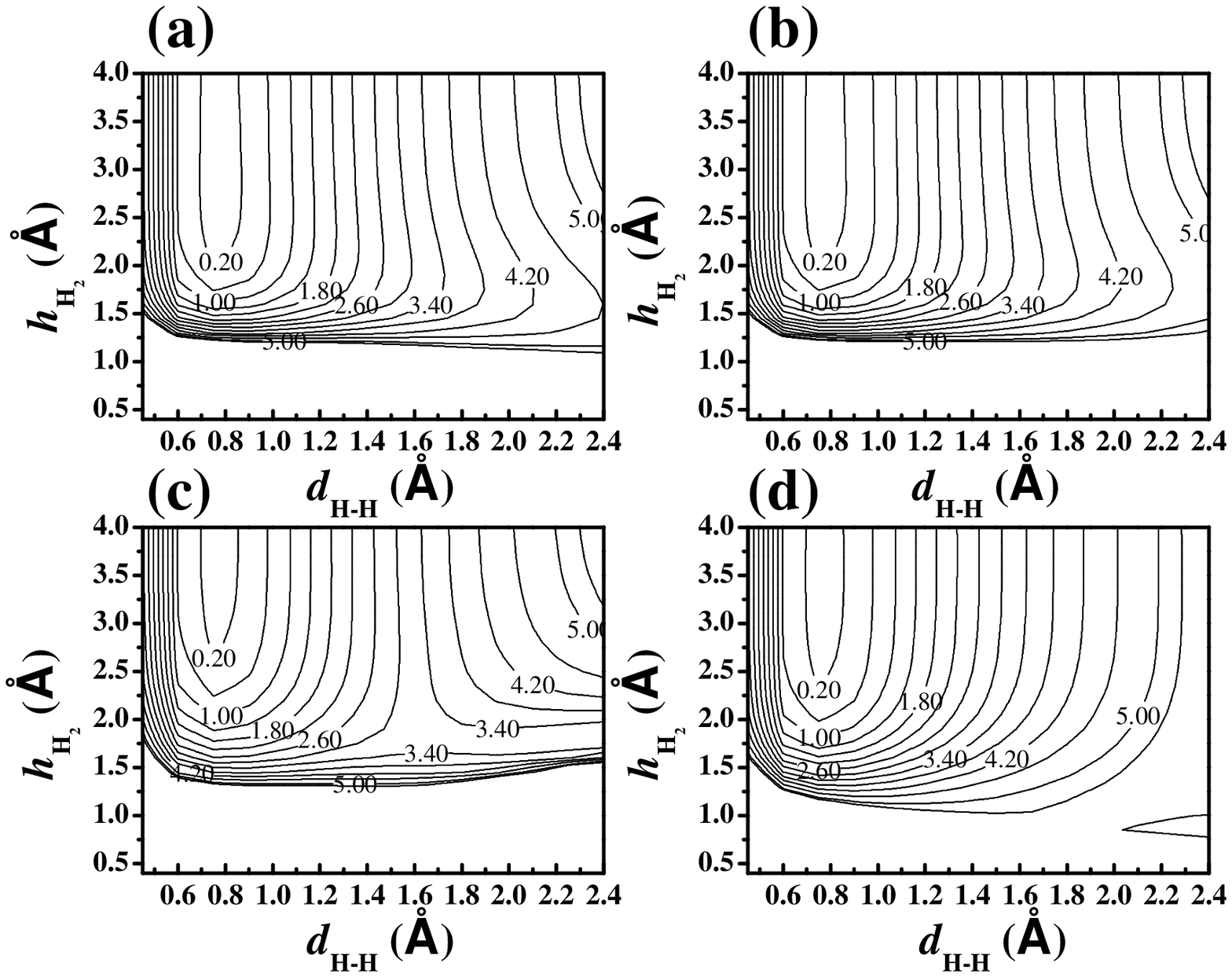}
\caption{\label{fig:fig4}}
\end{figure}
\clearpage
\begin{figure}
\includegraphics[width=1.0\textwidth]{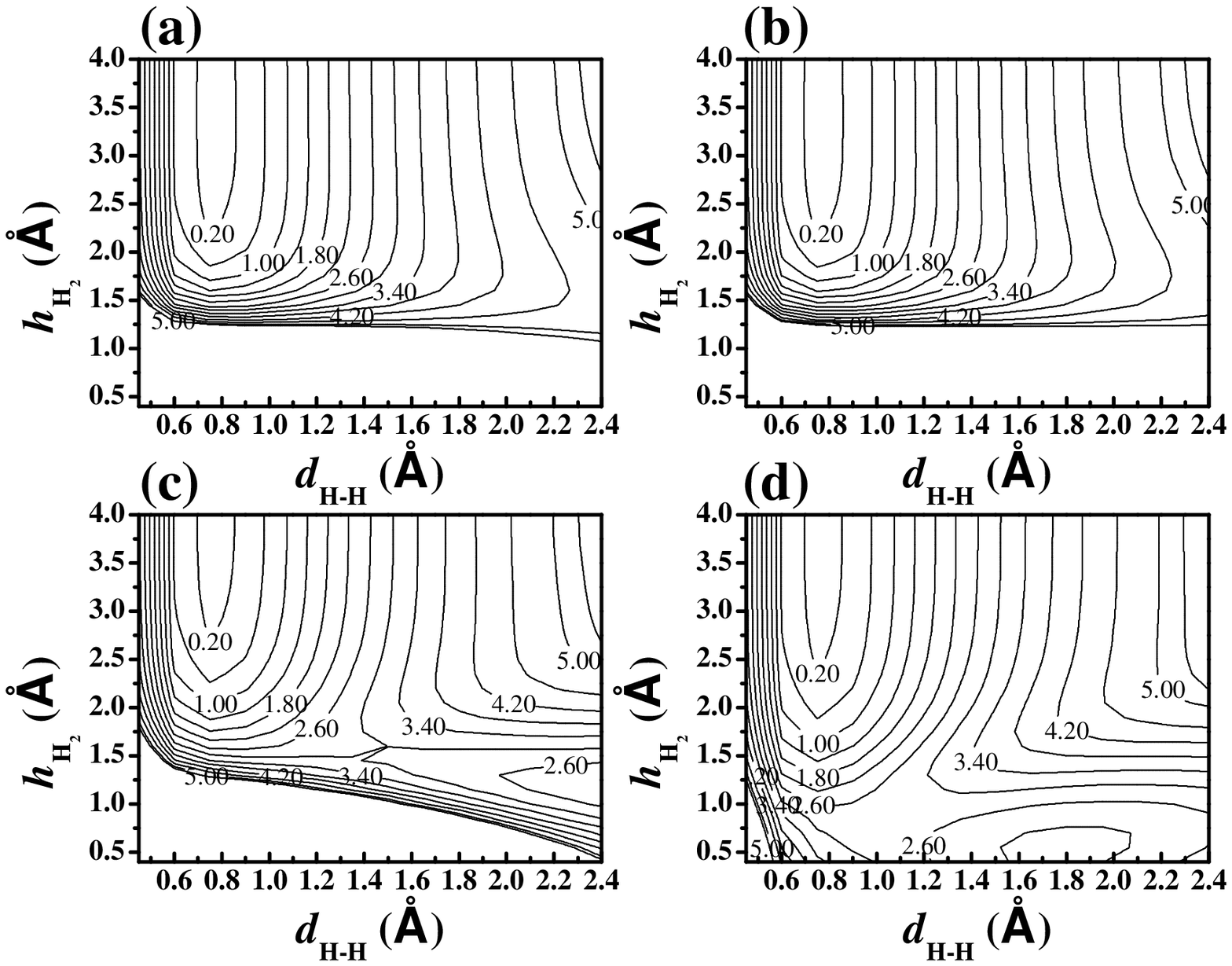}
\caption{\label{fig:fig5}}
\end{figure}
\end{document}